\documentclass[twocolumn,superscriptaddress,aps]{revtex4-1}
\usepackage{graphicx,epsfig,subfig,dcolumn,bm,mathrsfs,amsmath,amsthm,amssymb}
\begin{document}

\title{Computer simulations of single particles in external electric fields}

\author{Jiajia Zhou}
\email[]{zhou@uni-mainz.de}
\affiliation{School of Chemistry \& Enviroment, Center of Soft Matter Physics and its Application, Beihang University, Xueyuan Road 37, Beijing 100191, China}
\affiliation{Institut f\"ur Physik, Johannes Gutenberg-Universit\"at Mainz,\\
  Staudingerweg 9, D55099 Mainz, Germany} 
\author{Friederike Schmid}
\email[]{friederike.schmid@uni-mainz.de}
\affiliation{Institut f\"ur Physik, Johannes Gutenberg-Universit\"at Mainz,\\
  Staudingerweg 9, D55099 Mainz, Germany} 

\date{\today}

\begin{abstract}

Applying electric fields is an attractive way to control and manipulate single
particles or molecules, e.g., in lab-on-a-chip devices. However, the response
of nanosize objects in electrolyte solution to external fields is far from
trivial. It is the result of a variety of dynamical processes taking place in the ion
cloud surrounding charged particles and in the bulk electrolyte, and it is
governed by an intricate interplay of electrostatic and hydrodynamic
interactions.  Already systems composed of one single particle in electrolyte
solution exhibit a complex dynamical behaviour.  In this review, we discuss
recent coarse-grained simulations that have been performed to obtain a
molecular-level understanding of the dynamic and dielectric response of single
particles and single macromolecules to external electric fields. We address
both the response of charged particles to constant fields (DC fields), which
can be characterized by an electrophoretic mobility, and the dielectric
response of both uncharged and charged particles to alternating fields (AC
fields), which is described by a complex polarizability.  Furthermore, we give
a brief survey of simulation algorithms and highlight some recent developments.

\end{abstract}

\maketitle

\section{Introduction}
\label{sec:introduction}


Dispersions of nanoparticles in electrolyte fluids are ubiquitous in everyday
life. Prominent examples are proteins and DNA in aqueous environment. The
study of such systems is not only of technological interest for the development
of advanced materials, but also of fundamental interest for our understanding
of life science and biophysics. These materials have a large inherent
complexity.  Already the simplest system contains at least three components:
large macromolecules or particles (solutes), small electrolyte ions, and
solvent molecules. The interactions between the components include short-range
interactions, such as excluded-volume and van der Waals interactions, and
long-range interactions, such as the electrostatic and hydrodynamic
interactions. The interplay of these interactions determines the equilibrium
and dynamic properties of the system. Due to its multi-component nature, a
wealth of parameters can be used to control material properties, e.g., the
surface charge density, the salt concentration, the dielectric constant of the
solvent. External perturbations can be used to manipulate the behaviour of the
system. Since charges are involved, electric fields are particularly efficient.


In this review, we will focus on two important systems: Colloidal dispersions 
and polyelectrolyte solutions. Colloidal particles are solid objects with 
sizes ranging from a few nanometers to micrometers \cite{RSS, Dhont}.  
In electrolyte solution, an electric double layer forms near the solid/liquid 
interface and plays an important role in determining the dynamics.
Polyelectrolytes are charged polymers, with DNA and protein being the most
prominent examples \cite{Barrat1996, Dobrynin2005}.  In comparison to colloids,
which have no internal degree of freedom, polyelectrolyte chains can change 
their conformation from a coil to a compact globule, and to extended rodlike 
structures. Counterions can bind to the polyelectrolyte backbone in the case
of strongly charged chains, or form a diffusive layer. In both systems, it 
is important to consider the charged particle and its surrounding electric 
double layer as a whole. 


Computer simulations have become a widely accepted approach to studying
complex systems, complementing the well-established theoretical and
experimental methods. For charged particles in electrolyte solutions, the main
difficulty lies in the multi-scale nature of the system.  Taking colloidal
particles as an example, the length scales range from the size of the water
molecules (around $10^{-10}$m), to the size of the colloidal particle in the
order of $10^{-6}$m. Likewise, the dynamics of the system also involves
processes with characteristic time scales spanning several orders of magnitude.
For example, the characteristic diffusion time (the time to diffuse along one
molecule/particle diameter) for small salt ions and large colloids is of the
order of 10 picoseconds and seconds, respectively. In principle, one can use
molecular dynamics simulation with atomistic detail \cite{Yeh2004}, but even 
with current computer resources, the accessible time and length scales are 
still limited. One possible solution is to use coarse-grained simulations, 
which allow one to access larger length and time scales at the expense of
losing fine scale details.


In this paper, we review recent coarse-grained simulation studies of charged
colloids and polyelectrolyte chains in electrolyte solution. We mostly focus on
the case of one single particle and its response to a weak and spatially
homogeneous, albeit possibly time-dependent external electric field (linear
response regime). We discuss the physical mechanism behind various dynamic
processes driven by the external electric fields.  Since there is a vast amount
of theoretical and experimental literature on this subject, we shall not be
comprehensive, but we will focus on presenting simulations that illustrate
important physical mechanisms. We apologize in advance that our reference list
is far from complete.


This review is organized as follows: We start with a brief discussion of
simulation techniques in Section~\ref{sec:method}.  Then we turn to reviewing
simulation studies of particles and polyelectrolyte chains under constant 
electric fields (DC fields) and alternating electric fields (AC fields) 
in Section~\ref{sec:dc} and \ref{sec:ac}, respectively. We conclude in
Section~\ref{sec:summary} with a brief summary and perspective.

\section{Simulation Methods}
\label{sec:method}


When studying the dynamics of a system that includes large particles (colloids
or polyelectrolytes) and small solvents/ions, one needs to consider both the
electrostatic and hydrodynamic interactions.  From a simulation point of view,
modeling such a system is a challenging task, because both the electrostatic
and hydrodynamic interactions are long-range.  Taken separately, each
interaction has been well studied in the literature \cite{AllenTildesley,
FrenkelSmit, Slater2009, Pagonabarraga2010, Smiatek2012, Cisneros2014}.
Methods can be grouped into the categories of ``implicit'' and ``explicit''
methods, \cite{Pagonabarraga2010, Smiatek2012}, depending on whether the small
components (solvents and ions) are simulated explicitly or replaced by
effective interactions between large components. In this review, we shall focus
on explicit methods. The inclusion of small ions and solvent particles requires
more computational resources, but in many cases, this sacrifice is necessary
for a proper treatment of the dynamics. For example, implicit methods that
treat hydrodynamic interactions, e.g., at the level of an Oseen mobility matrix,
are not capable of accounting for the effect of finite Reynolds numbers, and
difficult to combine with complex boundary conditions. On the other hand,
implicit methods that replace the effect of small ions by screened
electrostatic potentials do not capture the multitude of complex dynamical
processes involved in the electrophoretic or dielectrophoretic response of a
particle to an electric field. In the following, we first mention some
classical approaches which are commonly used in mesoscopic simulations, and
then highlight a few new developments.

One popular method for treating Coulomb interactions is Ewald summation, which
applies to point charges with periodic boundary condition \cite{Ewald1921}.
The idea is to split the interaction into two contributions: one is
short-ranged and has a cutoff, the other one is a smooth function which can be
calculated efficiently in Fourier space.  After optimization, Ewald summation
scales as $O(\mathcal{N}^{3/2})$, where $\mathcal{N}$ is the total number of
point charges.  Several methods improve the scaling to $O(\mathcal{N}\ln
\mathcal{N})$ by computing the Fourier part on a grid using fast Fourier
transform.  Notable examples are Particle-Particle-Particle Mesh (P3M)
\cite{HockneyEastwood, Deserno1998, Deserno1998a}, Particle Mesh Ewald (PME)
\cite{Darden1993}, and Smooth Particle Mesh Ewald (SPME) \cite{Essmann1995}.
Methods that scale linearly $O(\mathcal{N})$ with the system size have also
been proposed in the literature, such as the Fast Multipole Method
\cite{Greengard1987}, and methods based on the Maggs approach \cite{Maggs2002,
Rottler2004, Pasichnyk2004}. These methods require an expensive computational
overhead or have a large prefactor in the linear scaling; thus Ewald-based
methods are still the most common choice in mesoscopic simulations.  Some of
the methods mentioned here have been combined into a parallel library ScaFaCoS
\cite{Arnold2013, ScaFaCoS}, which is freely available.

Mesoscale methods for simulating fluid dynamics are based on one simple
observation: As long as one is mostly interested in hydrodynamic effects, the
solvent dynamics can be replaced by an artificial dynamics as long as the
relevant conservation laws (mass, charge, momentum, etc.) are satisfied.
Therefore, one can design simple fluid models that can be simulated at low
computational cost.  Popular examples in the literature include the Lattice
Boltzmann (LB) method, Dissipative Particle Dynamics (DPD), and Multi-Particle
Collision Dynamics (MPCD). DPD is a particle-based method that uses soft
potentials and pair-wise interactions \cite{Hoogerbrugge1992, Espanol1995}.  LB
is a lattice-based method which solves a linearized Boltzmann equation in a
fully discretized fashion \cite{Succi, Duenweg2009}.  In MPCD, the interactions
between particles are performed by sorting particles in cells and followed by
local operations \cite{Malevanets1999, Kapral2008, Gompper2009}. Besides these
mesoscopic simulation methods, one can also solve the Navier-Stokes equations
numerically, for example, in Fluid Particle Dynamics \cite{Tanaka2000} and in
the Smoothed Profile method \cite{Nakayama2005}.

Once one settles on solvers for the electrostatic and hydrodynamic equations,
the next step is to choose a simulation model for the large particle.  In the
case of polyelectrolytes, a common choice is the bead-spring model, which
represents a polyelectrolyte as a chain of consecutive charged beads connected
by springs.  Large rigid particles, such as micrometer-sized colloids, can be
introduced through appropriate boundary conditions, e.g., no-slip boundaries
for the hydrodynamics and hard impenetrable boundaries for solvent and ion
particles. An alternative approach suitable for smaller colloids is the
so-called ``raspberry model'', which represents the colloid by a shell made of
of surface beads. The shape of the shell is maintained either by springs
\cite{Lobaskin2004} or by fixing the bead position with respect to the colloid
center \cite{Chatterji2005}.  

The computational expenses for the electrostatic and hydrodynamic computations
are not equivalent. At high salt concentrations, a comparison
\cite{Smiatek2009} has shown that the computational time is dominated by the
costs of treating the charges. Unfortunately, as mentioned above, implicit
models for small ions are not suitable for dynamic studies. We recently
proposed an efficient algorithm that overcomes the bottleneck caused by the
explicit charges \cite{2015_condiff}. It makes a compromise between computing
efficiency and taking full consideration of correlations at all scales.  The
evolution of the ionic concentration is computed using Brownian pseudo
particles \cite{Szymczak2003}, which is very fast because pseudo particles have
no direct pair interactions. In this approach, one chooses a coarse-grained
length scale. Short-range correlations between small ions on length scales
shorter than the coarse-grained length are neglected, but long-range
charge-charge correlations can be retained.  The full method consists of the
pseudo-ion solver for electrolytes, DPD for the fluid and a P3M Coulomb solver.
Simulations of electro-osmosis showed that the computer time required for the
electrostatic calculations can be reduced to about half the time required
for treating the fluid. Moreover, since the number of pseudo particles can be
chosen independently of the number of ions, it is independent of the ion
concentration. Therefore, the proposed method is particularly suited for
electrolyte solutions at high salt concentrations.

Another issue in electrostatic simulations is the dielectric constant, which is
often assumed to be homogeneous throughout the simulation box. In reality, the
dielectric permittivity in the water and in the nanoparticles differ
significantly.  The difficulty to include dielectric contrast in molecular
dynamics lies in the calculation of induced polarization charge at the
interfaces, which has to be done self-consistently in every time-step.
Recently, a number of authors have addressed this problem \cite{Tyagi2010,
Jadhao2012, Barros2014}. For example, Barros \emph{et al.} \cite{Barros2014}
proposed to use a Generalized Minimum Residual method \cite{Saad1986} to
calculate the distribution of surface bound charges on nanoparticles and
applied this approach to study the self-assembly of binary colloids.  They
observed an unexpected string structure formation due to the dielectric effect
\cite{Barros2014a}.  Boundary methods such as that sketched above also open up
the possibility to study phenomena associated with induced-charge
electrokinetics \cite{Bazant2004}. However, they cannot be used to simulate
media with a smoothly varying dielectric constant $\epsilon(\mathbf{r})$. Such
systems can be treated using the recently proposed Maxwell Equations Molecular
Dynamics (MEMD) method \cite{Rottler2004,Pasichnyk2004}, which is based on the
Maggs approach \cite{Maggs2002} and solves the Maxwell equations by propagating
a set of virtual auxiliary fields on a local scale. A recent review on methods
to deal with dielectric contrasts can be found in Ref.\ \cite{Arnold2013}.

\section{Particles in Constant Electric Fields}
\label{sec:dc}

Let us now consider an object (molecule or colloid) with a characteristic size
of $a$ suspended in an electrolyte solution. The object can acquire surface
charges by several means, either by the dissociation of protons or by the
selective adsorption of ions from the aqueous solution. Due to the
electrostatic interaction, it is surrounded by a so-called ``electric double
layer'' of oppositely charged ions. Some of these may bind strongly to the
surface (Stern layer), and their main effect is to reduce the effective
surface charge.  The others form the ``diffuse layer'', which is characterized
by a constant turnover of weakly bound, mobile ions.  The thickness of the
diffuse layer results from the competition of osmotic pressure and
electrostatic interactions and is typically of the order of the Debye screening
length $\kappa^{-1}$.  Both $\kappa^{-1}$ and the particle extension, $a$,
represent important characteristic lengths of the system.  When an external
electric field $\mathbf{E}$ is applied, a positively charged particle starts to
move in the direction of the electric field, and the ions in the diffuse layer
experience a force in the opposite direction. Thus the particle experiences a
friction force exerted by the electrolyte fluid which prevents its movement. In
the stationary case, the electric driving force and the friction force balance
each other and the object moves with a constant velocity $\mathbf{v}$.  The
ratio between the terminal velocity $\mathbf{v}$ and the applied electric field
$\mathbf{E}$ is called electrophoretic mobility $\mu$, 
\begin{equation}
  \mathbf{v} = \mu \mathbf{E}.
\end{equation}
In general, the electrophoretic mobility is a tensor, but we shall
here restrict the discussion to isotropic objects, in which case the
mobility is a scalar quantity.

Before we proceed to presenting simulations, we briefly recapitulate the main 
contributions to the friction force involved in 
electrophoresis \cite{Lyklema_vol2} 

\begin{enumerate}

\item The viscous force $\mathbf{F}_{\mathrm{vis}}$ exerted by the fluid.  The
magnitude of this force is given by the Stokes friction $F_{\mathrm{vis}} = -
6\pi\eta a v$, where $\eta$ is the shear viscosity of the fluid. 

\item The electrophoretic retardation force $\mathbf{F}_{\mathrm{ret}}$ due to
the movement of the counterion cloud.  Since the counterions have the opposite
charge, they move in the opposite direction of the central object.  Ideally,
the counterions compensate the charged object and the whole system is charge
neutral.  Therefore, the electrostatic force on the counterions exactly cancels
the driving force on the object, but to which extent this force is transferred
to the central object is complicated. 

\item The polarization force $\mathbf{F}_{\mathrm{pol}}$ due to the distortion
of the counterion cloud. As an example, let us consider a spherical colloid.
The charge centers of the colloid and its surrounding counterions coincide at
zero external field. If an external field is applied, the two centers are
displaced slightly as the counterion cloud is distorted, which induces a dipole
moment.  The central object experiences an extra electric force due to this
distortion. 

\end{enumerate}

\subsection{Electrophoresis of Colloidal Particles}
\label{sec:dc_colloid}

In general, the electrophoretic mobility of colloids depends on several
parameters, e.g., the colloidal size $a$, the Debye screening length
$\kappa^{-1}$, and the surface potential at the plane of shear, termed the
$\zeta$-potential. In simulation studies, it is sometimes easier to prescribe
the surface charge density $\sigma$. Unfortunately, the relation between
$\sigma$ and $\zeta$ is not simple \cite{Doane2011}; analytic formulas only
exist for simple geometries (planes and cylinders).  The electrophoretic
mobility has a simple form only for special cases such as weakly charged
colloids ($\zeta/k_BT \ll 1$).  In this limit, the counterion distribution can
be treated using the Debye-H{\"u}ckel approximation.  Two well-known results
are the H{\"u}ckel and Smoluchowski formulae:  If the electric double layer is
thick ($\kappa a \ll 1$), one can neglect the retardation and polarization
forces, and write the electrophoretic mobility as \cite{Hueckel1924}
\begin{equation} \label{eq:mu_Hueckel} \mu = \frac{2}{3} \frac{ \varepsilon }{
\eta } \zeta \quad (\kappa a \ll 1), \end{equation} where $\varepsilon$ is the
permittivity of the solution.  This formula is derived from the balance between
the electric driving force and Stokes friction.

In the opposite limit of thin electric double layers, one obtains the
Smoluchowski formula \cite{Smoluchowski1917}
\begin{equation}
  \label{eq:mu_Smoluchowski}
  \mu =  \frac{ \varepsilon }{ \eta } \zeta \quad (\kappa a \gg 1). 
\end{equation}
In this case, the retardation force is accounted for but the polarization force
is neglected.
The situation is different if the surface charge is kept constant. 
In this case, the surface potential vanishes for large $\kappa a$, resulting a zero mobility.

For intermediate values of $\kappa a$, Henry derived an analytic formula of the
form \cite{Henry1931}
\begin{equation}
  \label{eq:mu_Henry}
  \mu = f(\kappa a) \frac{ \varepsilon }{ \eta } \zeta
\end{equation}
with a scaling function $f(\kappa a)$ that interpolates between the
H{\"u}ckel and Smoluchowski result in their corresponding limits: 
$f(\kappa a) \to 1$ for $\kappa a \ll 1$ and $f(\kappa a) \to 2/3$ for $\kappa a \gg 1$.

The assumptions and approximations made in deriving these simple formulae are:

\begin{itemize}

\item 
The distortion of the counterion cloud is assumed to be negligible,
and the contribution from the polarization force $\mathbf{F}_{\mathrm{pol}}$ 
is omitted.

\item The distribution of salt ions and counterions is treated at the
mean-field level using a continuum approximation. 
In equilibrium, this corresponds to the Poisson-Boltzmann approximation. 
Short-range correlations between charged species are neglected. 

\item The colloidal particle is assumed to be weakly charged ($e\zeta/k_B T <
1$).  This assumption justifies the use of the Debye-H{\"u}ckel approximation,
and the simplification of the Poisson-Boltzmann equation to a linear equation
facilitates the derivation of analytic expressions. In this limit, the
counterion distribution near the charged surface decays exponentially with
the characteristic length $\kappa^{-1}$.  

\end{itemize}

We shall now discuss how theoretical considerations and numerical calculations
in the last few decades have helped to probe phenomena which are beyond 
those approximations.

\subsubsection{Methods Based on the Electrokinetic Equations}
\label{sec:dc_ek}

A general framework to study the dynamic phenomenon in electrolyte solutions is
provided by the electrokinetic equations \cite{RSS}.  This involves a change in
the point of view: Instead of calculating the friction force on the charged
colloids, one writes down explicitly a set of partial differential equations
which governs the dynamics of electrolyte solutions at the level of a continuum
theory. The electrokinetic equations have three basic constituents: The Poisson
equation for the electrostatic potential, the Navier-Stokes equations
characterizing the fluid flow, and the Nernst-Planck equation, which is
basically a convection-diffusion equation describing the time evolution of
ionic concentration profiles. Since thermal fluctuations and the discrete
character of charges are neglected in this continuum approach, it corresponds
to a mean-field approximation and reproduces the Poisson-Boltzmann
results at equilibrium. All components of the friction force are
implicitly included in this approach.

\begin{figure}[htp]
  \centering
  \includegraphics[width=1.0\columnwidth]{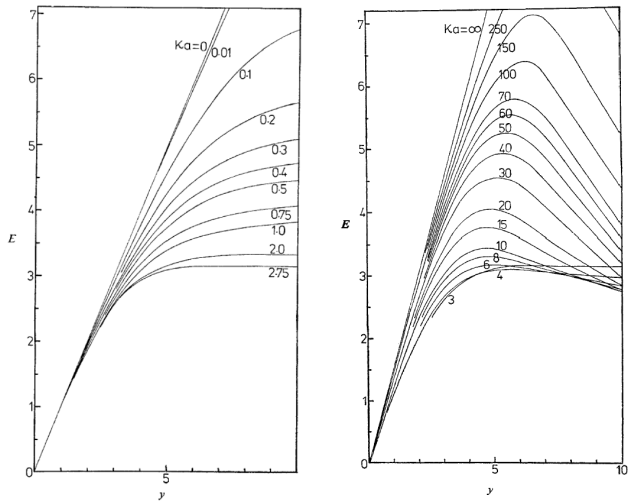} 
  \caption{
Reduced mobility $E=(6\pi\eta e/ \varepsilon k_BT) \mu$ as a function of the
reduced zeta potential $y=(e/k_BT) \zeta$. Left: The mobility is a monotonic
function for thick electric double layers $\kappa a < 3$.  Right: A mobility
maximum appears for thin electric double layers, $\kappa a > 3$. Reprinted from
Ref.~\cite{OBrien1978} with permission from The Royal Society of Chemistry.}
  \label{fig:1}
\end{figure}

In an early seminal paper \cite{OBrien1978}, O'Brien and White computed the
electrophoretic mobility of a single sphere in an infinite domain by
numerically solving the electrokinetic equations for weak external fields
(linear regime). Their results differed qualitatively from the previous
predictions in the case of the thin electric double layer: Instead of being a
monotonically increasing function of the zeta-potential $\zeta$, as predicted
by Eq.~(\ref{eq:mu_Smoluchowski}) , the mobility has a maximum for values of
$\kappa a > 3$.  This is shown in Fig.~\ref{fig:1}, right.  The non-monotonic
behaviour results from the fact that different competing factors contributing
to electrophoresis scale differently with $\zeta$: The electrostatic driving
force is proportional to $\zeta$, while the friction force associated
with the distorted counterion cloud is proportional to $\zeta^2$. 

Based on the same electrokinetic model, Schmitz and D{\"u}nweg
\cite{Schmitz2012} recently developed a lattice-based approach  to solve the
linearized electrokinetic equations. Their strategy is divide-and-conquer: They
divided the original equations into parts, and developed different numerical
solvers for each equation.  They then combined different solvers to compute the
original equations using an iteration procedure. Their method is different to
O'Brien and White in two aspects: One is the usage of periodic boundary
condition, which allows to study the effect of colloid concentration.  The
second difference is that no assumption is made about the colloid shape, and
one can in principle study colloidal particle with irregular shape.

It is also possible to start with the full electrokinetic equations instead of
the linearized version.  The three equations belong to different categories of
partial differential equations, and there exist efficient solvers for each
equation separately.  Based on the Lattice Boltzmann method, Giupponi and
Pagonabarraga studied the electrophoretic mobility of charged colloids
\cite{Giupponi2011}. The Nernst-Planck equation was solved using a discrete
method \cite{Capuani2004}. The results match those of O'Brien and White well
in the case of small zeta-potentials, but deviate when the zeta-potential is
large. In this regime, the method used by O'Brien and White suffers from
numerical problems \cite{OBrien1978}, which may explain the discrepancy.  In
particular, Giupponi and Pagonabarraga found that a mobility maximum exists
for all salt concentrations.  They further investigated the effect of
diffusivity of the small ions. 

Kim \emph{et al.} took a different approach to the fluid by solving the
Navier-Stokes equations directly \cite{Kim2006}. The difficulty lies in the
proper and efficient treatment of the moving boundary of the colloids. They
devised the Smoothed Profile method, which replaces the shape boundary by a
smooth interface with finite thickness \cite{Nakayama2005}.  Their results for
the electrophoretic mobility showed good agreement with O'Brien and White for
thick electric double layer ($\kappa a = 0.5$).  The efficiency of their
numerical method permits the simulation of many colloids in one simulation box.
They further investigated the mobility dependence on the colloid concentration
and compared their results with the theoretical prediction of Ohshima
\cite{Ohshima1997}.  At low colloid concentration, the agreement between the
simulation and theory is good. Deviations become noticeable when the electric
double layers from different colloids overlap. 

\subsubsection{Particle-based Simulations}
\label{sec:dc_particle}

All studies based on the electrokinetic equations neglect short-range
correlations, as the small ions are treated in terms of ionic concentration
fields. To remove this approximation, one must simulate the small ions as
particles with excluded volume interactions that carry discrete charges.
Combined with mesoscopic methods for the fluid, one obtains a simulation
scheme which accounts for correlations between charged species. One drawback
is an increase in computational time which makes the simulation of large
colloids very difficult.

Lobaskin \emph{et al.} studied the electrophoresis of charged colloids in
electrolytes containing only counterions or with very low salt content
\cite{Lobaskin2007, Duenweg2008}. In their approach, the fluid is simulated
using the LB method, and the colloidal particle is modeled using the raspberry
model \cite{Lobaskin2004}.  Using a dimensional analysis, they demonstrated
that the reduced electrophoretic mobility, defined by $\mu_{\rm red} =
\mu/\mu_{\rm H}$ ($\mu_{\rm H}$ is the H{\"u}ckel result,
Eq.~\ref{eq:mu_Hueckel}), depends only on two dimensionless parameters $Z_{\rm
eff} l_B/a$ and $\kappa a$.  Here $Z_{\rm eff}$ is the effective
charge of the colloid, which differs from the bare colloid charge $Z$ by the
amount of strongly adsorbed counterions, $l_B=e^2/4\pi\varepsilon k_BT$ is
the Bjerrum length, and the inverse screening length $\kappa$ is defined as
\begin{equation}
  \label{eq:kappa_ci}
  \kappa^2 = 4\pi l_B (c_i + c_s), 
\end{equation}
where $c_i=Z/V$ is the counterion concentration and $c_s$ is the ionic
concentration due to the salt ions. Based on the dimensional analysis and
Eq.~(\ref{eq:kappa_ci}), one can map the mobility of colloids in salt-free fluids
(the number of counterions is set by the colloid concentration) to that in
electrolytes containing a small amount of salt at the same value of $\kappa a$.
This correspondence is confirmed by both the simulations and experiments
\cite{Lobaskin2007}.

A similar model was used by Chatterji and Horbach in a series of studies
of highly charged colloids \cite{Chatterji2005, Chatterji2007, Chatterji2010}.
They varied the colloid charge density instead of the zeta potential because
the former can be controlled more easily in simulations.  Upon increasing the
charge density, the electrophoretic mobility was found to initially increase,
then reach a maximum, and decrease again. These simulation results are in
accordance with the predictions for the thin electric double layer case
(Fig.~\ref{fig:1}, right). 

Chatterji and Horbach also considered systems containing divalent counterions,
in which they found the electrolyte mobility to be reversed at high charge
density, indicating an overcompensation of the surface charge by the
multivalent counterions. This overcharging phenomenon can be explained by ion
correlations \cite{Quesada-Perez2003} and has also been observed
experimentally.  In experiments, however, the effect is sometimes larger than
predicted by numerical simulations, suggesting that it might be enforced by ion
specific attractive forces \cite{Semenov2013}. Recent simulations by
Raafatnia \emph{et al.} have shown that overcharging may even occur in
electrolytes containing only monovalent ions if the colloids are coated by a
suitable organic layer \cite{Raafatnia2014a, Raafatnia2015}.

As mentioned above, treating colloids with sizes much larger than the ion size
is difficult with explicit ion models.  Raafatnia \emph{et al.} have developed
a hybrid approach to studying the electrophoresis of large colloids
\cite{Raafatnia2014}.  They first measured the zeta-potential of a flat surface
using simulations with explicit microions.  The measured zeta-potential was
then used as an input in the electrokinetic equations to calculate the
electrophoretic mobility.  When compared with experiments, the scheme worked
well for monovalent and divalent salt solutions. In the case of trivalent
salt, they needed to introduce an attractive interaction in order to 
reproduce the mobility reversal which is observed experimentally.  

\subsection{Polyelectrolytes}
\label{sec:dc_pe}

We turn to discussing polyelectrolyte electrophoresis. Compared
to colloid electrophoresis, there are several differences:

\begin{itemize}

\item Colloidal particles are usually at least one order of magnitude larger
than the small molecules (solvents and salt ions). The separation of length
scales makes it possible to describe large colloids in terms of boundary
conditions and motivates the use of continuum approaches. In the case of
polyelectrolytes, the size of the monomer units is comparable to that of small
molecules and the application of boundary methods becomes questionable. Even
though the zeta potential is sometimes used to parameterize experimental
results, its physical meaning is not always obvious. In simulations, an
explicit treatment of small ions is often more appropriate than the use of
continuum theories.

\item Whereas colloidal particles are rigid objects, polyelectrolyte chains 
can assume many conformations. The conformational dynamics is strongly
influenced by the electrostatic and hydrodynamic interactions between monomers.
Therefore, a full treatment of both interactions is necessary. Implicit
methods have been designed to circumvent this requirement for stationary
situations \cite{Hickey2010,Hickey2012}, but in general, one has to be
careful when dealing with non-stationary states. The conformational
flexibility complicates studies of electrophoretic mobility.

\item If polyelectrolytes are highly charged, the strong electrostatic
interaction attracts counterions to the proximity of the chain backbone. This
phenomenon, called Manning condensation \cite{Manning1969}, also influences 
the electrophoretic mobility.

\item In the case of colloidal particles, the size and the surface charge
density can be adjusted independently. For polyelectrolytes, this is usually
not the case.  The control parameters are the chain length $N$ and the charge
fraction.  We shall restrict ourselves to strongly charged chain, where each
monomer carries a unit charge.  The size of a polyelectrolyte chain is
characterized by the radius of gyration $R_g$, which depends on the chain
length. Therefore, the electrophoretic mobility is usually represented as a
function of the chain length.

\end{itemize}

\subsubsection{Free Solution Electrophoresis}

When the size of polyelectrolyte chain is much smaller than the Debye screening
length, $\kappa R_g \ll 1$, one may invoke the H{\"u}ckel picture for colloidal
particles and consider only the electric driving force and the viscous
friction.  The driving force of the external field is proportional to the chain
length, while the viscous force from the fluid also increases with the chain
length, but to a smaller extent due to the hydrodynamic interactions. Monomers
experience hydrodynamic drag and shield each other from the exposure to the
outer fluid, which effectively reduces the total Stokes friction.
Therefore, for short chains, the mobility increases with the chain length.

This behaviour was indeed observed in simulations and experiments.  Grass and
Holm modeled the polyelectrolyte as a chain of charged beads connected by
finitely extensible nonlinear elastic bonds \cite{Grass2008}.  The fluid was
simulated using the LB method, and the Coulomb interaction was calculated using
the P3M method. They computed the electrophoretic mobility from equilibrium
simulations using the corresponding Green-Kubo relation. Later they also
directly measured the terminal velocity in an external electric field
\cite{Grass2009, Grass2010}, and found that the results agree in the weak-field
regime. Frank and Winkler used a similar model for the polyelectrolyte chain,
but simulated the fluid using MPCD \cite{Frank2008}.  Both studies 
emphasize the importance of hydrodynamic interaction, as complementary
Langevin simulations show a decrease in the mobility with increasing
chain length.

\begin{figure}[htp]
  \centering
  \includegraphics[width=1.0\columnwidth]{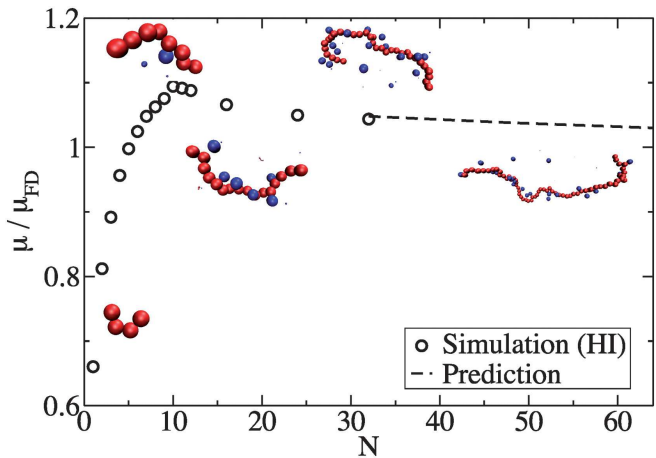} 
  \caption{Reduced electrophoretic mobility $\mu$ of a polyelectrolyte chain as
a function of the chain length $N$. The mobility is scaled with the
free-draining limit $\mu_{\rm FD}$. Reprinted from Ref.~\cite{Grass2009} with
permission from The Royal Society of Chemistry.}
  \label{fig:2}
\end{figure}

One peculiar observation in these simulations is the presence of a mobility
maximum at chain lengths $N \sim 10$ (see Fig.~\ref{fig:2}). This was
attributed to a reduction of the effective charge of the polyelectrolyte chain,
which reduces the electric driving force. As the chain becomes longer,
the polyelectrolyte assumes a rod-like conformation due to the mutual repulsion
of monomers, and Manning condensation sets in.  The accumulation of counterions
on the backbone reduces the effective charge of the chain, thereby reducing the
electric driving force.  Frank and Winkler measured the number of condensed
counterions in their simulation using a distance criterion \cite{Frank2008}.
They found that as the chain becomes longer, the ratio between the condensed
counterion to the total charge of polyelectrolyte chain approaches $1-1/\xi$,
where $\xi$ is the Manning parameter.  Grass and Holm applied different
estimators for the effective charge, with similar results \cite{Grass2009}.
However, the exact point at which the saturation is reached is still unclear,
because the polyelectrolyte chain is flexible, whereas the theory of Manning
condensation applies for rigid rods. 

When the chain length increases further, the electrophoretic mobility reaches a
constant value.  This so-called ``free-draining'' limit has been well studied
in the literature \cite{Manning1981, Barrat1996}. The fact that the mobility
does not depend on the length of the polymers prevents the separation of long
polyelectrolyte chains by free solution electrophoresis.  The physics behind
the plateau in mobility is two-fold: On the one hand, the effective charge per
monomer becomes a constant, once the Manning condensation sets in.  Therefore,
the electric driving force is proportional to the chain length.  On the other
hand, the effective friction force opposing the electric force also scales
linearly with the chain length once the polyelectrolyte size becomes larger
than the Debye screening length. This is because a long chain can be viewed as
a chain of charge neutral blobs with the blob size being the Debye length. When
applying an external electric field, the chain experiences a force, but the
surrounding counterions experience a force in the opposite direction, such that
the sum of all forces on charged particles in a blob is zero. Therefore, no
flow is induced, the hydrodynamic interactions associated with the electric
force are screened \cite{Kekre2010}, and the resulting electrophoretic mobility
does not depend on the chain length.

\subsubsection{Effect of Adding Salt}

The hydrodynamic screening effect discussed above becomes even more pronounced
in the presence of salt. Salt plays a dual role in polyelectrolyte
electrophoresis, because it screens both the electrostatic and the hydrodynamic
interactions. The screened electrostatic potential decays exponentially with a
characteristic length $\kappa^{-1}$. The screened hydrodynamic flow profile
still features a long-range $1/r^3$ decay, which is due to the vectorial nature
of the velocity field \cite{Long2001}.  In free solution, the screening length
for hydrodynamics $\kappa_{\mathrm{H}}^{-1}$ coincides with the Debye screening
length $\kappa^{-1}$.  Therefore, varying the salt concentration induces a
complicated interplay between the electrostatic and hydrodynamic interactions.
In Ref.~\cite{Grass2010}, Grass and Holm examined the effect of salt on the
electrophoretic mobility of short polyelectrolyte chains.  The increase in the
salt concentration reduces the screening length $\kappa_{\mathrm{H}}^{-1}$ and
therefore reduces the initial shielding of hydrodynamics.  As a result, the
electrophoretic mobility becomes almost length-independent at high salt
concentration.  Fischer \emph{et al.} \cite{Fischer2008} found that the
counterion mobility changes sign as a function of salt concentration.  At low
salt concentration, the counterions condense to the polyelectrolyte backbone
and the hydrodynamic drag forces the counterions to move together with the
polyelectrolyte chain. At high salt concentrations, however, the screening of
hydrodynamics decouples the motion of counterions and polyelectrolyte, and
counterions move in the opposite direction of the polyelectrolyte. 

\subsubsection{Effect of Confinement}

Another way of decoupling the screening lengths for electrostatics and
hydrodynamics is to use confinement. The screening length for hydrodynamics is
given by either the characteristic size of the confinement or the Debye length,
depending on which is smaller.  Hickey \emph{et al.} investigated the
electrophoretic stretching of a polyelectrolyte chain confined between parallel
plates \cite{Hickey2014}.  They found that the hydrodynamic interaction is
screened by the confining walls for strongly confined chains.  Their study
focused on uncharged walls.  In the case of charged wall, the electroosmotic
flow induced by the counterions from the charged surface can reverse the
movement of the polyelectrolyte chain: a positively charged polyelectrolyte
chain can move in the opposite direction of the external field if confined by
negatively charged walls \cite{Smiatek2010, Smiatek2011}.

\section{Particles in Alternating Electric Fields}
\label{sec:ac}

Alternating electric fields (AC fields) provide another attractive 
tool for manipulating particles. Compared to constant electric fields 
(DC fields), one has several advantages:

\begin{itemize}

\item In homogeneous AC fields, the displacement of the charged particle 
averages to zero. This is in contrast to the DC case, where particles may
travel over long distances during the time of an experiment. Since charged
particles are mostly stationary, electrodes can be placed close to each
other. Therefore, one can produce large value of the electric field, which 
is in general difficult in the DC case.

\item Using AC fields, one can avoid the accumulation of charged species
on electrodes.

\item Unlike DC fields, AC fields 
do not generate constant electro-osmotic flows.

\item Apart from the amplitude, one can also tune the frequency and the phase
of AC fields. Since dynamical processes in the system
can take place on different time scales, a time-dependent perturbation can
probe the dynamics on selective time scales. 

\item Finally, AC fields can also be used to manipulate uncharged particles.
This is because they can induce polarization charges both in charged and
uncharged particles, and the polarized particles can then be further
manipulated by applying electric field gradients.

\end{itemize}

\subsection{Colloidal Particles}
\label{sec:ac_colloids}

The dielectric response to AC fields can be characterized by the
polarizability $\alpha(\omega)$, where $\omega$ is the field frequency.  In
analogy to the electrophoretic mobility, the polarizability is defined as the
ratio between the induced dipole moment and the external field,
\begin{equation}
  \label{eq:alpha_def}
  \mathbf{p} = \alpha(\omega) \mathbf{E}.
\end{equation}
The dipole moment has contributions from the colloidal particle and its
surrounding electric double layer.  In general, the polarizability depends on
both the frequency and the amplitude of the external field. In weak fields,
however, linear response applies and the polarizability does not depend on
the field strength. In strong electric fields, nonlinear effects may 
become important.

One important application of AC fields is dielectrophoresis \cite{Pohl, Jones}.
In dielectrophoresis, the time-averaged force on a particle has the form $\Re
\{ \alpha(\omega) \} \nabla |\mathbf{E}|^2$, where $\Re \{\alpha(\omega)\}$ is
the real part of the complex polarizability.  Under the influence of a
position-dependent AC field, the particle is driven along the direction of the
field gradient \cite{Salonen2005, Salonen2007, ZhaoHui2011}.  The magnitude of
the force depends on the polarizability of the particle, which permits
separation of colloidal particles and biomacromolecules \cite{Regtmeier2007}.
In a spatially homogeneous setup, AC fields can be used to control the
self-assembly of many particles \cite{Grzelczak2010}.  The control is realized
by tuning the induced dipole-dipole interaction between particles. 
The most important quantity in this context is the complex
polarizability of the charged particle. Here we shall focus on studies of
single colloidal particle and its polarizability under a homogeneous field.

\subsubsection{Maxwell-Wagner Theory and Electrokinetic Theory}

The calculation of the induced dipole moment of a single spherical particle
immersed in the medium provided by an electrolyte solution is a classical
problem in electrodynamics, known as Maxwell-Wagner mechanism of dielectric
dispersion \cite{Maxwell1954, Wagner1914}.  The complex polarizability
$\alpha(\omega)$ has the form
\begin{equation}
  \label{eq:alpha_mw}
  \alpha(\omega) = 4\pi \varepsilon_{\rm m} a^3 \mathcal{K}(\omega) = 4\pi \varepsilon_{\rm m} a^3 \frac{ \varepsilon_{\rm p}^* - \varepsilon_{\rm m}^* }{ \varepsilon_{\rm p}^* + 2 \varepsilon_{\rm m}^*},
\end{equation}
where $\mathcal{K}(\omega)$ is the Clausius-Mossotti factor.  The system is
characterized by the complex dielectric functions $\varepsilon_{\rm p,m}^*
= \varepsilon_{\rm p,m} + K_{\rm p,m}/i\omega$, where $\varepsilon$ is the
permittivity and $K$ the conductivity, and the subscript (${\rm p,m}$)
labels the particle or the medium.  From the form of the complex dielectric
constant, one notices that the permittivity contribution dominates at high
frequency ($\omega \rightarrow \infty$), while the conducting properties are
more important at low frequency ($\omega \rightarrow 0$).  The frequency
separating these two limits is the inverse of the Maxwell-Wagner relaxation
time
\begin{equation}
  \label{eq:tmw}
  \tau_{\mathrm{mw}} = \frac{\varepsilon_{\rm p} + 2\varepsilon_{\rm m}}{K_{\rm p} + 2K_{\rm m}}.
\end{equation}

For charged colloids, the electric double layer introduces another contribution
to the dipole moment.  O'Konski proposed a surface conductance term to account
for the effect of electric double layer \cite{OKonski1960}, and the same
mechanism was also extended to the study of ellipsoidal colloids
\cite{Saville2000}.  The Maxwell-Wagner-O'Konski theory has a simple analytic
formulation which gives qualitatively correct predictions, but it suffers from
one main drawback:  Since the theory is entirely formulated in terms of
macroscopic properties, such as the conductivities and permittivities, the
polarization charges are taken to be localized at the particle/medium
interface, and their spatial distribution is entirely neglected. This
simplification is only justified for thin electric double layers and in the
high-frequency regime.

Experiments revealed another dispersion in the low-frequency regime, which
cannot be explained by the Maxwell-Wagner-O'Konski theory. This low-frequency
dispersion, often called alpha-relaxation, results from the distortion of the
electric double layer.  To properly treat the electric double layer, one can
resort to the electrokinetic equations. Dukhin and Shilov made the further
assumption that the electric double layer is at local equilibrium with the
surrounding bulk solution \cite{DukhinShilov}, which is only valid in the
low-frequency regime.  They derived an analytic theory for the thin electric
double layer case, which correctly predicts the low-frequency dispersion. The
theory has been extended to asymmetric electrolytes 
\cite{Hinch1984, Chassagne2003, Grosse2009a, Grosse2009b} and to
aspherical colloids \cite{Chassagne2008}.

In situations that involve thick electric double layer and the whole frequency
spectrum, one can solve the full electrokinetic equations using a variety of
numerical methods.  DeLacey and White \cite{DeLacey1981} extended the method of
O'Brien and White \cite{OBrien1978} to AC fields and presented numerical
solutions for the polarizability of a single spherical colloid \cite{DeLacey1981}.
The electrokinetic equations are linearized in term of the external field, thus
the calculation assumes weak fields. The method was further refined by
Mangelsdorf \emph{et al.} \cite{Mangelsdorf1992,Mangelsdorf1997} and Zhou
\emph{et al.} \cite{Zhou2005}. Hill \emph{et al.} developed an alternative
numerical scheme which overcomes a numerical instability in the high-frequency
regime \cite{Hill2003}, and extended the method to polymer-grafted
particles \cite{Hill2003a, Hill2005}.  The effect of the colloid concentration
can also be incorporated by performing the calculations in a sphere, whose
volume is equal to the inverse of the colloid number density
\cite{Carrique2008a, Carrique2008b, Roa2012}.  Shih \emph{et al.} applied the
Smoothed Profile method \cite{Kim2006} to study the response of charged
colloids to AC fields and reproduced both the Maxwell-Wagner relaxation and the
alpha-relaxation \cite{Shih2014,Shih2015}.

The numerical methods were also extended to rodlike particles.  Zhao
investigated the case of long parallel rod with two spherical caps on both ends
\cite{ZhaoHui2010}.  The calculation showed that the transition frequency in the
low-frequency dispersion is reduced when the rod becomes longer, and do not
change once the rod length reaches a certain limit. Dhont and Kang 
developed theories based on electrokinetic equations to study two special
cases: Very weakly charged rods where the dipole moment is mainly induced
by the obstacle of the solid colloids \cite{Dhont2010}, and strongly charged 
rods where condensed counterions fully compensate the bare rod
charge \cite{Dhont2011}.

\subsubsection{Simulations with Explicit Ions}

\begin{figure}[htp]
  \centering
  \includegraphics[width=1.0\columnwidth]{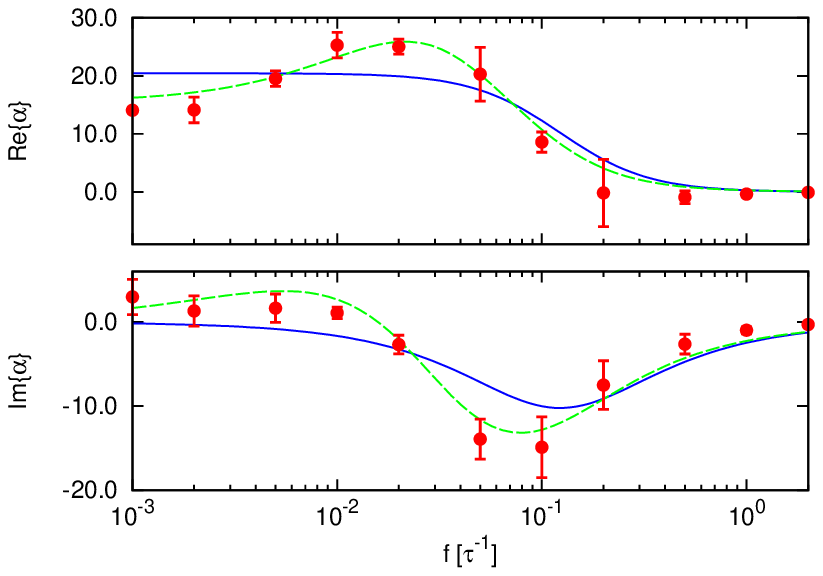}
  \caption{Real and imaginary part of the complex polarizability
$\alpha(\omega)$ of a charged particle as a function of the frequency. The
field strength is set in the linear region. The points with err-bars are
simulation results. The solid lines give the prediction from the
Maxwell-Wagner-O'Konski theory. The dashed lines are numerical solutions to the
electrokinetic equations~\cite{Hill2005}. Reprinted from
Ref.~\cite{2013_ac_review}.}
   \label{fig:3}
\end{figure}

The present authors have used particle-based method to investigate the
dielectric response of a charged nanometer-sized colloid \cite{2012_ac,
2013_q0,2013_response, 2013_ac_review}.  The fluid dynamics is simulated using
DPD, where small ions are included explicitly as charged beads with
excluded-volume interactions. The large colloid is modeled using the raspberry
model. We systematically investigated the complex polarizability
$\alpha\{\omega\}$ as a function of the frequency, shown in Fig.~\ref{fig:3}. 

Let us first discuss the situation in the low-frequency regime. In the absence
of an external field, the system including the charged colloid and its
surrounding electric double layer has spherical symmetry. When an external
field is applied, the colloid (positively charged) moves in the direction of
the electric field, while the counterions move in the opposite direction. This
creates a dipole moment, which points  in the same direction than the external
field. It also distorts the electric double layer, compressing it at the
front end and expanding it at the back end, and thus creates a concentration
gradient from back to front. In addition, the colloid acts as a moving obstacle
for the salt ions in solution which are also driven by the electric field.
Negatively charged ions accumulate at the front end of the colloid, which
combined with the extra counterions results in a further increase of the salt
concentration there, whereas the concentration at the back is further reduced.
The concentration gradient induces a diffusive migration of the salt molecules
from the front to the back. This concentration-induced effect reduces the
dipole moment created by the external field.  However, since it is a
second-order effect caused by the field-induced dipole, the net dipole moment
still points in the direction of the external field.  This can be seen from
Fig.~\ref{fig:3}, where the real part ${\rm Re}\{\alpha\}$ has a positive value
at low frequency.

The diffusion of the salt over the distance of colloid diameter requires time,
which can be estimated by
\begin{equation}
  \label{eq:tc}
  \tau_{\rm c} \sim \frac{(2a)^2}{D_i},
\end{equation}
where $D_i$ is the diffusion constant of the small ions.  If the field
frequency increases beyond $1/\tau_{\rm c}$, the external field oscillates so
fast that the diffusion process cannot follow. Therefore, the
concentration-induced process is suppressed at frequencies $f>1/\tau_{\rm c}$,
which effectively increases the dipole moment.  This is the origin of
the low-frequency dispersion and explains the slight increase of
$\Re\{\alpha\}$ in Fig.~\ref{fig:3}.

When the frequency is further increased, the external field eventually 
oscillates so fast that the ion cloud cannot respond. Thus, both ${\rm
Re}\{\alpha\}$ and ${\rm Im}\{\alpha\}$ drop to zero.  The characteristic
time corresponds to the time required for the small ions to diffuse over the
distance of Debye length \begin{equation} \tau_{\rm mw} \sim
\frac{\kappa^{-2}}{D_i} = \frac{\varepsilon_{\rm m}}{K_{\rm m}}.
\end{equation} The last equality applies for our simulation parameters
($\varepsilon_{\rm m}=\varepsilon_{\rm p}$ and $K_{\rm p}=0$), which differs
from Eq.~(\ref{eq:tmw}) only by a prefactor.

The simulation results can be compared with the theoretical and numerical
predictions. The Maxwell-Wagner-O'Konski theory gives the solid curves in
Fig.~\ref{fig:3}, which are only in qualitative agreement with the simulation.
The theory predicts roughly the correct transition frequency in the
high-frequency regime, but misses the low-frequency dispersion. The numerical
solution of the electrokinetic equations is shown with dashed lines. It is in
quantitative agreement with the simulation results.

We have also investigated the dielectric response of uncharged colloids to
external AC fields \cite{2013_q0}. The Maxwell-Wagner theory, Eq.~(\ref{eq:alpha_mw}), predicts a complex polarizability, simply due to the fact
that the medium is conducting and the particle is not. Physically, the particle
acts as an obstacle for the flow of charges induced by the applied field, such
that charges of opposite sign accumulate at both ends. Since uncharged
particles have no electric double layer, one does not expect additional
contributions. Indeed, the simulation data were found to be in almost perfect
agreement with the predictions of the Maxwell-Wagner theory. Fine details of
the charge distributions can be understood quantitatively within the 
electrokinetic theory developed by Dhont and Kang \cite{Dhont2010}.

\subsection{Polyelectrolytes}
\label{sec:ac_pe}

Simulating the  polyelectrolyte response to AC fields is a challenging
task and only a few researchers have tackled this problem.  Most studies
focused on the conformational change of a single polyelectrolyte chain when a
strong field is applied. For polyelectrolyte chains that are sufficiently long,
or in the case of high salt concentration, hydrodynamic interactions are
strongly screened. Therefore, most simulation studies used a Langevin
thermostat which neglects the hydrodynamics, but includes small ions explicitly
in order to account for the Manning condensation.

Liu \emph{et al.} studied the unfolding and collapse of a flexible
polyelectrolyte under a sinusoidal electric field \cite{LiuHongjun2010}.  They
first measured the critical field strength at which the chain undergoes a
transition to the fully extended state.  Their results confirmed a theoretical
prediction by Netz \cite{Netz2003a,Netz2003c}: The critical field scales as
$E_{\rm crit} \sim N^{x}$, where $x=-1/2$ for collapsed chains and
$x={-3\nu/2}$ for non-collapsed chains with the Flory exponent $\nu$.  They
also estimated the relaxation time by measuring the correlation time of the
end-to-end distance.  For certain parameters of the AC field, the
polyelectrolyte chain exhibit a stretch-collapse cycle. The simulations
indicated such a behaviour occurs under two conditions: (1) the field strength
is larger than the critical strength and (2) the frequency is comparable or
less than inverse of the chain relaxation time.  Hsiao \emph{et al.} examined a
similar system, but in trivalent salt solutions under a square-wave electric
field \cite{HsiaoPai-Yi2011}.  They implemented a different estimator for the
relaxation time.  Since they use a square wave, the dipole moment exhibits
an exponential decay after the electric field reverses its sign.  They used the
characteristic time for the exponential decay to estimate the relaxation time
and reached the same conclusions than Liu \emph{et al.} regarding the 
requirement for chain stretching. 

\section{Conclusions}
\label{sec:summary}

Strategies to control particles with electric fields have attracted
considerable attention in recent years. Since the particles respond to the
external electric fields on relatively short time scales and in an often fully
reversible way, using electric fields provides an attractive approach to control
the position of individual particle or the structure of particle assemblies.
In this review, we have discussed recent simulation studies of charged colloids
and polyelectrolyte chains under external electric field.  These studies have
identified two important length scales: the Debye screening length $\kappa^{-1}$
and the size of the particles (the radius $a$ of colloidal particles or the
radius of gyration $R_g$ of polyelectrolytes).  In the case of AC fields, the
frequency-dependent response is determined by the diffusion times ($\tau_{\rm
mw}$ and $\tau_c$) associated with those two length scales.  In most
situations, the delicate interplay between the electrostatic and the
hydrodynamic interactions plays an important role in determining the dynamics.

Although many studies have addressed the behaviour of particles in external
electric fields, many open questions still remain.
For example, we have only discussed the behaviour of spherical
particles with homogeneous surface properties. Removing this constraint opens
up the whole new territory of anisotropic particles \cite{Glotzer2007}.
Colloidal particles can have heterogeneous surface charge distributions
\cite{Long1998, Bianchi2014}, or varying hydrodynamic slip \cite{Swan2008,
Khair2009, ZhaoHui2010a}.  These novel types of colloidal particles and their
response to electric fields still remain to be explored.

We have focused on flexible polyelectrolyte chains, while most interesting
polymers are semiflexible, for example, double-stranded DNA.  Understanding how
semiflexible polyelectrolytes move under external fields is not only important
from a scientific point of view, but can also help advancing molecular biology
and medicine research \cite{Viovy2000, Dorfman2010}.  Regarding the
conformational changes of polyelectrolyte chains under external fields, we have
only touched one aspect of the story: the chain stretching due to the external
fields.  Experiments have shown that it is possible to \emph{collapse}
polyelectrolyte chains under DC or AC fields \cite{TangJing2011,
ZhouChunda2011}.  The mechanism for this surprising behaviour is still under 
debate, and input from theories and simulations would certainly help to
understand such phenomenon.

Furthermore, colloidal particles and polyelectrolytes chain can be seen as
two extremes of particles with varying rigidity.  There exists a zoo of
so-called soft-particles that interpolate between the solid colloids and
flexible polyelectrolytes \cite{Ohshima2009, Cametti2011}.  Notable examples
are polymer-grafted colloids, star polyelectrolytes, and microgels.
Simulations can help to understand how to control and manipulate those particle
using electric fields.

In this review, we have focused on the case of dilute suspensions.  In a dense
suspension or at low salt concentration, the electric double layers of
different particles may overlap. This introduces another length scale,
the average distance between charged particles. External fields may modify
the effective interaction among particles, which in turn influences the 
self-assembly process in dense suspensions. This opens new possibilities
for manipulating particles, which may be used for directed self-assembly.
Simulations may be useful to guide the design of experimental protocols 
for making novel interesting nonequilibrium structures.


\begin{acknowledgments} 

We would like to thank many coworkers and colleagues who have helped to shape
our understanding in this subject, in particular Burkhard D{\"u}nweg, Christian
Holm, Vladimir Lobaskin, Stefan Medina, Taras Molotilin, Roman Schmitz, Jens
Smiatek, Olga Vinogradova, and Zhen-Gang Wang. This work was supported by the
German Science Foundation (DFG) within SFB 1066 and SFB TRR146.  

\end{acknowledgments}

\bibliography{codef}

\end{document}